\documentstyle[preprint,tighten,prc,aps,epsfig]{revtex}

\begin{document}
\title{$\eta{N}$ Final State Interaction in 
incoherent Photoproduction of $\eta$-mesons from the Deuteron near
Threshold.} 

\author{A. Sibirtsev$^1$, S. Schneider$^1$, Ch. Elster$^{1,2}$, 
J. Haidenbauer$^1$, S. Krewald$^{1}$ and J. Speth$^1$}
\address{
$^1$Institut f\"ur Kernphysik, Forschungszentrum J\"ulich,
D-52425 J\"ulich \\
$^2$Institute of Nuclear and Particle Physics,
Ohio University, Athens, OH 45701 }

\date{\today}
\maketitle

\begin{abstract}
An analysis of incoherent photoproduction of $\eta$ mesons 
off the deuteron for photon energies from threshold to 800 MeV 
is presented. The dominant 
contribution, the $\gamma$N-$\eta$N amplitude, is described within an
isobar model. Effects of the final state interactions in the
$NN$ as well as the ${\eta}N$ systems are included employing models
derived within the meson-exchange approach. It is found that their 
consideration is important. Specifically, due to an interference 
effect the influence of the $\eta N$ final state interaction is 
enhanced in the reaction $\gamma d \to np \eta$ close to threshold.  
\end{abstract}

\pacs{13.60.Le; 13.75.-n; 13.75.Cs; 14.40.Aq, 14.20.Gk}

\section{Introduction}
\label{sec1}
Already for a long time 
the determination of the low energy properties of the $\eta{N}$ interaction 
has been a challenging subject in meson-nucleon physics. 
In fact, the actual size of the $\eta{N}$ scattering length is still an
open question and there is  substantial disagreement between 
different theoretical predictions. 

The  knowledge about the low energy behavior of the
 $\eta{N}$ interaction is extracted  either from the analysis of  resonance 
models of the reactions
$\pi{N}{\to}{\eta}N$ and $\gamma{N}{\to}{\eta}N$ \cite{Bhalerao,Bennhold,Wilkin1,Faldt}
or from coupled channel $K$-matrix~\cite{Sauerman1,Sauerman2,Green2,Green3} or 
$T$-matrix~\cite{Arima,Batinic1,Abaev,Batinic2} approaches, which 
in addition  include the available information on $\pi{N}$ scattering.
The current  predictions of  the imaginary
part of the $\eta{N}$ scattering length vary 
between 0.16 and 0.37~fm, while the variation of the real part 
ranges from 0.25 to 1.05~fm. Calculations based on 
an effective Lagrangians approach~\cite{Kaiser,Kaiser2,Kaiser3,Nieves} 
tend to provide scattering lengths on the lower end of the range  
given above whereas a result from a chiral heavy baryon expansion~\cite{Krippa} 
lies roughly in the middle of this range.

In principle, the $\eta{N}$ interaction could be deduced~\cite{Bauer} from 
$\eta$-meson production in nuclear reactions like $\gamma$-nucleus, 
$\pi$-nucleus, $p$-nucleus, or 
heavy ion collisions. However, in those cases the extracted $\eta{N}$ 
parameters should be considered as being relevant only for in-medium 
$\eta$-meson scattering and might not represent the vacuum 
parameters~\cite{Effenberger}. Thus, the study of $\eta$-meson production 
in few-body systems is certainly more promising with regard to that,
specifically because for such reactions a wealth of rather accurate
data has become available over the last few years. E.g., 
the $\eta{N}$ scattering length might be inferred from 
experiments on $pd{\to}^3$He$ \eta$~\cite{Mayer1,Betigeri,Klimala} or
$np{\to}d\eta$~\cite{Calen1,Calen2} by applying models which
relate the measured $^3$He $\eta$ and $d\eta$ scattering 
length with the one for $\eta{N}$. In this context it should be noted
that both, the $pd{\to}^3$He $\eta$
and $pn{\to}d\eta$ measurements, indicate a sizable final state interaction
(FSI) due to the $\eta$-meson, and thus may be sensitive to the model 
employed.

Of course it should be expected that also the reactions
$pp{\to}pp\eta$~\cite{Chiavassa,Calen3,Bergold,Smyrski} 
and $\gamma{d}{\to}np\eta$ data~\cite{Krusche1} show sensitivity 
to the $\eta{N}$ FSI for energies near the production threshold. 
However, it is still an open question whether the analysis of these 
reactions will allow to pin down the $\eta{N}$ scattering
length in a model independent way. 
Therefore it is important to investigate a larger class of $\eta$-production 
reactions and see to what extent predicted values for the $\eta$N scattering 
length are compatible with each other.

Here we investigate the effect of the $\eta{N}$ FSI in incoherent 
photoproduction of $\eta$-mesons from the deuteron near threshold.
We extend a previous study~\cite{Our}, where the
$\gamma{d}{\to}np\eta$ total and differential cross sections 
were calculated by considering the impulse approximation as well as the
neutron-proton ($np$) FSI. In that work it had turned out that the
$np$ FSI can account for a large part of the experimentally observed 
enhancement of the production cross section near the threshold. 
However, some discrepancies to the data still remain, specifically 
at very small excess energies. This discrepancy could be a signal of the
$\eta N$ FSI and therefore we want to address this question in the 
present paper. 
In our previous study we also established that the
effect of the $np$ FSI does not depend on the specific model for the
nuclear force, since the contributions of intermediate states with large momenta are
suppressed through the integration over the deuteron wave function. This feature,
specific to the photoproduction reaction on the deuteron, 
is definitely of advantage if one wants to study the 
sensitivity of the cross section near threshold to the $\eta$N interaction.

For the $\eta N$ FSI we utilize a microscopic model 
developed by the J\"ulich group~\cite{Schutz99,Krehl},
which is a coupled channels model for $\pi N$ scattering that includes 
the $\eta N$ channel and quantitatively describes the $\pi N$  phase shifts
and inelasticity parameters in both isospin channels for partial waves up to
$J = {3\over 2}$ and pion-nucleon center-of-mass (c.m.) 
energies up to 1.9 GeV \cite{Krehl}. 
Specifically, it provides a realistic description of the quantities 
relevant for the present investigation, namely  the $S_{11}$ $\pi N$ phase 
shift and  the $\pi N \to \eta N$ transition cross section. 
In Section \ref{sec2} we give a short introduction to this 
model, discuss its predictions for   
the $\eta{N}$ interaction and estimate relevant 
uncertainties of the model. 
In Section \ref{sec3} we describe the details of our treatment 
of the $np$ and $\eta N$ FSI in the reaction $\gamma{d}{\to}np\eta$.
Furthermore, we present our results and analyze the influence of
the $\eta N$ FSI in detail.  
Finally, in Section \ref{sec4}, we provide a short summary. 

\section{$\eta{N}$ interaction}
\label{sec2}
The basic information about the low energy behavior of any interaction is
given by the scattering length. Since the enhancement of the 
$\eta$ photoproduction cross section is only seen for very low ${\eta}$
momenta, one may assume that this enhancement is caused primarily by the 
low energy behavior of the $\eta{N}$ interaction, and thus by the physics
contained in the ${\eta}N$ scattering length $a_{\eta{N}}$. 
However, as already discussed in the Introduction, this quantity is not 
too well determined. A list of currently available values
for $a_{\eta{N}}$ is given in Table~\ref{tab1}
together with the method and reactions involved 
in the extraction of those values. This compilation\protect\footnote{We 
note that in status review~\cite{Green1} of $\eta{N}$ scattering length the 
results of Ref.~\protect\cite{Wilkin1} were misinterpreted as being 
extracted from $pd{\to}^3He\eta$ data. The $\eta{N}$ scattering 
length~\protect\cite{Wilkin1,Green1} actually comes from analysis of 
the $\pi^-p{\to}n\eta$ data.} 
clearly indicates that the present status of knowledge about the
$\eta{N}$ scattering length is far from being satisfactory.

In our investigation of the influence of the ${\eta}N$ 
FSI in the $\eta$ photoproduction reaction 
we employ a meson-exchange model developed 
for $\pi N$ scattering~\cite{Schutz99,Krehl}, which includes the 
$\pi{N}$- and $\eta{N}$-channels, and in addition reaction channels which can 
decay into two pions and a nucleon, namely $\sigma{N}$, $\rho{N}$ and 
$\pi\Delta$. Here $\sigma$ and $\rho$ are notations for the correlated exchange of 
two pions in the scalar-isoscalar and vector-isovector channels, respectively. 
The model describes the ${\pi}N$ phase shifts and 
inelasticities up to 1.9 GeV quantitatively, and also reproduces available data
for $\pi^-p{\to}\eta{n}$ differential cross sections. Thus, it is very
well suited for the investigation we want to carry out. In particular 
the model has the advantage 
that it allows us to calculate the $\eta N$ reaction amplitude for any
on- as well as off-shell momenta, as it is required for evaluation of the
${\eta}N$ FSI effects, without making additional assumptions.

In Fig.~\ref{diag0} the Feynman diagrams that define the 
quasi-potential for the $\pi{N}$-${\eta}N$ and the 
${\eta}N$-${\eta}N$ transitions are shown. These diagrams are the
relevant ones for the calculation of the $\eta N$ t-matrix, 
which  is derived within time-ordered 
perturbation theory in order to have a  well-defined 
off-shell behavior. Further details of the model can be found in Ref.~\cite{Krehl}. 
The $\pi{N}$-channel is coupled with the $\eta{N}$-channel by the 
$s$-channel exchanges of  the $N^*(1520)$, $N^*(1535)$, 
and $N^*(1650)$ baryonic resonances. The exchange of 
a nucleon in the $u$-channel is also taken into account. In addition, we 
allow for the exchange of an effective $a_0$-meson in the $t$-channel,
representing correlated $\pi\eta$ exchange \cite{Janssen}.

Within this model~\cite{Krehl}, the optimal fit to the 
$\pi{N}$ phase shifts and inelasticities  overestimates  the total 
$\pi^-p{\to}\eta{n}$ cross section by roughly 15$\%$. In order to reproduce the
$\pi^-p{\to}\eta{n}$ cross section, we have weakened the 
coupling of the effective $a_0$-meson at the $\pi\eta{a_0}$ vertex. 
This leads to an underestimation 
of the $\pi{N}$ inelasticities in the $S_{11}$ partial wave in 
the vicinity of the $N^*(1535)$ resonance. In Ref.~\cite{Feuster}, 
this problem is handled  by the introduction of an effective 
$\pi\pi{N}$ reaction channel, which has quantum numbers that 
cannot be interpreted by resonant two-pion states, however.
One may take the results of Ref.~\cite{Feuster} as a hint that
the approximate  treatment of the three-body dynamics of the 
$\pi\pi{N}$ channel by resonances is not adequate. In the context of this work here, 
we do not try to resolve this problem, but will 
consider the above mentioned change in the coupling of the $a_0$-meson as theoretical
uncertainty of our model and discuss its consequences on the size of the low energy
$\eta N$ parameters and the $\eta$ photoproduction cross section.

The real and imaginary parts of the scattering amplitude  in the
$\eta{N}$ partial wave amplitude
$S_{11}$ are shown in Fig.~\ref{maeta1} as function of 
the $\eta$-meson momentum $q$ given in the $\eta{N}$ c.m. system. 
The symbols represent the  scattering amplitude  obtained from 
the full calculation, while the solid line indicates the effective 
range expansion of the scattering  amplitude with the scattering 
length $a_{\eta N}$ and effective range $r_{\eta N}$ given by
\begin{eqnarray}
a_{\eta N}&=&0.42 + i 0.34 \,\,\, \mbox{fm}  \nonumber \\
r_{\eta N}&=&-2.0 + i 0.8 \,\,\,\,\, \mbox{fm}.
\label{ouref}
\end{eqnarray}
Here the low energy parameters are related to the on-shell ${\eta}N$
scattering amplitude $f(q)$ by
\begin{equation}
f(q) = \left[ \frac{1}{a_{\eta N}}+\frac{r_{\eta N}q^2}{2}-iq \right]^{-1},
\label{efra}
\end{equation}
and $a_{\eta N}{=}\lim_{q{\to}0}f(q)$. 
The relation between the ${\eta}N$ scattering amplitude  and 
$t_{\eta N}$-matrix is given in our normalization as 
\begin{equation}
f(q) = - \pi \frac{\sqrt{q^2+m_N^2}\sqrt{q^2+m_\eta^2}}
{\sqrt{q^2+m_N^2}+ \sqrt{q^2+m_\eta^2}}\,\, t_{\eta N}(q,q), 
\end{equation} 
where $m_N$ and $m_\eta$ are the nucleon and $\eta$-meson masses,
respectively. In order to show the dependence of the ${\eta}N$ 
scattering amplitude on the low energy parameters, 
the dashed line in Fig.~\ref{maeta1} gives the effective range expansion of
$f$ with the parameters of Ref.~\cite{Green3}, namely
\begin{eqnarray}
a_{\eta N}&=&0.75 + i 0.27 \,\,\,\, \,\mbox{fm}  \nonumber \\
r_{\eta N}&=&-1.5 - i 0.24 \,\,\, \mbox{fm}.
\label{greenef}
\end{eqnarray}
It is very clear that the two different scattering amplitudes exhibit a
substantial difference both in absolute value and in the momentum dependence
of the ${\eta}N$ scattering amplitude.

Table~\ref{tab1} indicates that the $\eta{N}$ interaction is frequently 
described in a resonance model approach. In order to demonstrate 
the strong interplay between resonant and non-resonant contributions  
in the case of the J\"ulich model we isolate the resonant piece and 
compare its amplitude to the one given by the full model.
The solid symbols in Fig.~\ref{maeta1s} indicate the real (dots) and 
imaginary (squares) part of the $\eta{N}$ $S_{11}$ partial wave
scattering amplitude calculated with the $N^\star$(1535) resonance 
alone. The solid lines represent the 
total $\eta{N}$ scattering amplitude given by the effective range 
expansion with the parameters of Eq.~(\ref{ouref}) and correspond to 
the solid lines in Fig.~\ref{maeta1}. This comparison 
clearly shows that the non-resonant contribution is quite large 
at small momenta $q{\le}300$~MeV/c, and thus the interplay between resonant and
non-resonant contributions is very important for the scattering length.

\section{The Reaction $\gamma{d}{\to}np\eta$.}
\label{sec3}
The lowest order contributions for the reaction $\gamma{d}{\to}np\eta$ 
are depicted in Fig.~\ref{diag1}. 
In our  previous study~\cite{Our} of this reaction we considered the
production of $\eta$ mesons via the dominant $S_{11}$(1535) 
resonance (impulse approximation), shown in Fig.~\ref{diag1}a, and the  
$np$ FSI as indicated in Fig.~\ref{diag1}b.
The explicit details of the model for the direct production 
are given in Ref.~\cite{Our}, and will only be briefly summarized here.
Using the impulse approximation (IA) the amplitude ${\cal M}_{IA}$ of
the reaction $\gamma{d}{\to}np{\eta}$ for given
spin $S$ and isospin $T$ of the final nucleons
can be written as
\begin{equation}
{\cal M}_{IA}{=}A^T(s_1)\phi(p_2){-}(-1)^{S+T}A^T(s_2)
\phi (p_1).
\end{equation}
Here $\phi(p_i)$ stands for the deuteron wave function, $p_i$ ($i=1,2)$ is the momentum 
of the proton or neutron in the deuteron rest frame, and $A^T$ denotes
the isoscalar or isovector $\eta$-meson photoproduction amplitude at
the squared invariant collision energy $s_N$ given by
\begin{equation}
s_N=s-m_N^2-2(E_\gamma+m_d)E_N+2\vec{k}_\gamma\cdot\vec{p}_i.
\end{equation}
The photon momentum is given by ${\vec k}_\gamma$. 
Details of the
photoproduction amplitude $A^T$ are described in Ref.\cite{Our}.

The result for the total cross section for the reaction
$\gamma{d}{\to}np\eta$ based on the
impulse approximation is shown as dotted line in
Fig.~\ref{maeta10_bl} in comparison to experiment~\cite{Krusche1}. 
The cross section is shown as a function
of the photon beam energy $E_\gamma$ (lower axis) as well as a function of
the excess energy (upper axis). Here the excess energy is defined as
\begin{equation}
\epsilon = \sqrt{s}-m_p-m_n-m_\eta,
\label{excess}
\end{equation}
with the invariant mass $s{=}m_d^2{+}2m_dE_\gamma$. 
The particle masses adopted in our
calculations are
$m_d$=1875.61339~MeV, $m_p$=938.27231~MeV, $m_n$=939.56563~MeV
and $m_\eta$=547.3~MeV.
We explicitly indicate the masses with the available 
accuracy, since the excess energy depends on those values,
and thus the accuracy is important for the analysis of the data
 especially very close to the reaction threshold.
The comparison with experiment shows that the impulse approximation
reproduces the data well for excess energies $\epsilon{\ge}50$~MeV,
however there is a clear underprediction of the data for energies closer
to threshold.

In the next order of the scattering expansion, the FSI
in the nucleon-nucleon (NN) system and the $\eta$-nucleon
system need to be considered. The effect of the FSI
in the NN channel is expected to be the stronger one, and
was already studied previously \cite{Our}, but for
completeness we repeat the essential ingredients here.
The amplitude describing the NN FSI can be written as
\begin{equation}
{\cal M}_{NN}=t_{NN}\, g_{NN}\, A^T.
\end{equation}
Here $t_{NN}$ denotes the half-shell $np$ scattering matrix and $g_{NN}$ the 
two nucleon propagator. Explicitly, the amplitude for the diagram
shown by  Fig.~\ref{diag1}b is given by
\begin{equation}
{\cal M}_{NN}=m_N\int dk\, k^2\, \frac{t_{NN}(q,k)\, A^T(s^\prime_N) \,
\phi(p^\prime_N)}{q^2-k^2+i\epsilon}, \label{3.4}
\end{equation}
with $q$ being the nucleon momentum in the final $np$ system and
\begin{equation}
\vec{p}^{\,\prime}_N=\vec{k}+\frac{\vec{k}_\gamma -\vec{p}_\eta}{2}.
\end{equation}
Here $p_\eta$ represents the $\eta$-meson momentum and 
$p^\prime_N{=}|\vec{p}^{\,\prime}_N|$.
The half-shell $np$ scattering matrix in the
$^1S_0$ and $^3S_1$ partial waves $t_{NN}(q,k)$ is obtained
at corresponding off-shell momenta $k$
from the CD-Bonn potential \cite{Machleidt1}, which describes the
$NN$ data base with a $\chi^2$/datum of about 1.  However, as already
pointed out in Ref.~\cite{Our}, all modern, high precision NN interaction 
would give the same result, since due to the integration over the
deuteron wave function in Eq.~({\ref{3.4}) possible off-shell differences
in the NN potentials are strongly suppressed.

The amplitude with the $\eta$-N FSI is graphically
represented in Fig.~\ref{diag1}c and given as
\begin{equation}
{\cal M}_{\eta N}=t_{\eta N}\, g_{\eta N}\, A^T,
\end{equation}
where $t_{\eta N}$ is the half-shell ${\eta}$N scattering matrix. The
$\eta$N propagator is indicated by $g_{\eta N}$. Explicitly, the amplitude
is written as
\begin{equation}
{\cal M}_{\eta N} = \frac{m_N m_\eta}{m_N+m_\eta}
\int dk\,  k^2  \, \frac{t_{\eta N}(q,k) \,
A^T(s^{\prime\prime}_N)\,   \phi (p^{\prime\prime}_N)}
{q^2-k^2+i\epsilon}.
\label{etanfsi}
\end{equation}
The $\eta$-meson momenta in the final and intermediate state of the
$\eta{N}$ system are given by $q$ and $k$, $\phi$ stands for the deuteron
s-wave, and  $t_{\eta N}(q,k)$ is the
half-shell $\eta{N}$ scattering matrix
in the $S_{11}$ partial wave, and
\begin{equation}
\vec{p}^{\,\prime\prime}_N= \vec{k} +
\frac{m_N \, (\vec{k}_\gamma-\vec{p}_N)}{m_N+m_\eta},
\end{equation}
where $\vec{p}_N$ is the momentum of the outgoing proton or neutron in 
the deuteron rest frame.

The total cross section $\gamma d \rightarrow np\eta$ including the
$np$ and $\eta N$ FSI in s-waves is displayed in
Fig.~\ref{maeta10_bl}.
The dashed line shows the result for the impulse approximation plus
$np$ FSI. 
As pointed out in Ref.~\cite{Our}, the consideration of the $np$ FSI alone
provides already an almost satisfactory description of the experimental
cross section for small excess energies. The additional consideration of
the $\eta$N FSI as given by the meson-baryon model described in Section
II leads to the solid line in Fig.~\ref{maeta10_bl}. As expected, the
$\eta$N FSI enhances the cross section very close to the reaction
threshold. However, for excess energies in the range of $10 \ge \epsilon
\ge 40$~MeV the presently available experimental values are slightly
overpredicted. We also observe that for excess energies larger than
40~MeV there is no effect of the $\eta$N FSI anymore. The same is true
for the $np$ FSI. 

It is illuminating to look at the difference in the relative strength of the two
different final state interactions
and their possible interference in our full calculation of the $\eta$
photoproduction cross section. For a more detailed insight, we plot in
Fig.~\ref{maeta3_1} the
ratio of calculations with the final state interactions included separately to the
calculation based on the impulse approximation alone. The dotted line in
Fig.~\ref{maeta3_1}
represents the calculation including only the $\eta N$ FSI, the dashed line only the
$NN$ FSI. From this, it is clear that the $NN$ FSI is the dominant one. The ratio of the
full calculation containing both, the $\eta N$ and $NN$ FSI, to the impulse
approximation
is given by the solid line. A comparison to the two other curves shows that the two
final state interactions interfere constructively at small excess energies,
which magnifies the effect of the relatively weak $\eta N$ FSI.

In order to study the effects of the $\eta$N interaction close to the
reaction threshold in more detail, we divide the experimental points
by our calculation containing the IA and the $np$ FSI, i.e. the result of
the dashed line in Fig.~\ref{maeta10_bl}. Now the enhancement 
in the cross section for the reaction $\gamma d \rightarrow np\eta$ 
close to the threshold becomes visible by a deviation from one for excess 
energies $\epsilon \leq 30$~MeV, cf. Fig.~\ref{maeta3_bl}. 
This enhancement, which is present in the data, is seen in
the full calculation containing the $\eta$N FSI as well. The solid line of
Fig.~\ref{maeta10_bl} corresponds to the lower bound of the shaded band in 
Fig.~\ref{maeta3_bl}. The $\eta$N interaction discussed in Section II
leads to a scattering length $a_{\eta N}=0.42 + i 0.34$~fm
(cf. Eq.~(\ref{ouref})).
Since this model leads to a slight overprediction of the $\pi^-p \to \eta n$
cross section we have produced a variant model that also fits the  data (cf.
the discussion in Section II). In the latter model the $\eta$N interaction 
is somewhat more attractive, leading to a scattering length of 
$a_{\eta N}{=}0.54{+}i0.18$. The corresponding 
total cross section for the reaction $\gamma d \rightarrow np\eta$
is given in Fig.~\ref{maeta3_bl} by the upper bound of the shaded area. 
Consequently, the shaded area in Fig.~\ref{maeta3_bl} can be seen 
as a reflection of the theoretical uncertainty in the $\eta N$ interaction
arising within the J\"ulich meson-baryon model.

Indeed, it is interesting to explore more thoroughly how strongly our 
calculation of the $\eta$N FSI depends on the specific properties of
the $\eta$N interaction. In order to study whether there is sensitivity 
to the off-shell behavior we carried out calculations where the $\eta$N 
amplitude is replaced by its effective range expansion of Eq.~(\ref{efra}), 
i.e. basically by the on-shell $\eta$N t-matrix. (Note that 
the quality of the effective range expansion as compared
to the full $\eta$N t-matrix is shown in Fig.~\ref{maeta1}.)
Surprisingly, we obtained  identical results
for the $\eta$ photoproduction cross section 
when using either the $\eta$N t-matrix from the effective range
expansion or the one from the full model.
This can be explained through the relatively good representation of 
the $\eta{N}$ scattering amplitude by the effective range expansion for 
quite a large momentum range as is illustrated in Figs.~\ref{maeta1}
as well as a very weak $k$-dependence of the half-shell 
$\eta{N}$ t-matrix, $t_{\eta N}(q,k)$. 
As an aside, we want to emphasize that an argument along the same line
with respect to the $np$ FSI and its effective range 
expansion is not valid, since the energy dependence as well as
the dependence on the off-shell momentum of the $NN$ t-matrix is much 
stronger. 

The next logical step is to see whether even more quantitative information about the
strength of the $\eta N$ interaction at low energies can be extracted. 
Since we found that the $\eta N$ amplitude obtained with the effective range
expansion of Eq.~(\ref{efra}) is numerically identical to the full calculation of the
amplitude, Eq.~(\ref{etanfsi}), 
we can explore the effect of different values for the
$\eta$N low energy parameters on the cross section of the
reaction $\gamma d \rightarrow np\eta$ close to threshold.
Our studies could establish
 that the calculations of the $\eta$-photoproduction cross section
are not sensitive to the effective range parameter
$r_{\eta N}$ of Eq.~(\ref{efra}),  leaving the only sensitivity to the
scattering length $a_{\eta N}$.
In Fig.~\ref{maeta3_bl} we display two calculations with scattering lengths 
$a_{\eta{N}}{=}0.25{+}i0.16$~fm~\cite{Bennhold} (dashed line) and
$a_{\eta{N}}{=}0.74{+}i0.27$~fm~\cite{Green2} (dash-dotted line).
The figure suggests that the presently available data 
for the total cross section for $\epsilon{\le}$40~MeV show a preference
for a smaller value of the $\eta{N}$ scattering length. 
Obviously more precise data are highly desirable to confirm these
indications. Since the calculations with different values for $a_{\eta N}$
involve differences in the real as well as imaginary part of the scattering 
length, we investigated which of the
two is mainly responsible for the effects shown in Fig.~\ref{maeta3_bl} and found
that the major contribution to the size of the $\eta N$ FSI in this reaction stems
from the interference of the real part of $a_{\eta N}$ with the $NN$ FSI.

\section{Summary}
\label{sec4}
We calculated the reaction $\gamma d \rightarrow np\eta$ including
the dominant contribution by the $S_{11}(1535)$ resonance and 
the final state interactions between all outgoing particles. 
Those final state interactions influence the cross section for inclusive
photoproduction of $\eta$ mesons only for excess energies of the $\eta$-meson
smaller than 40~MeV. At higher energies the cross section is solely
given by the impulse approximation.

The FSI between the outgoing
nucleons is essential to bring the calculated cross section into the
vicinity of the experimental values. Due to an interference 
effect the $\eta N$ final state interaction provides an
additional enhancement of the production cross section at energies 
close to threshold as required by the data. 
We found that the effect of the FSI
resulting from the $\eta N$ interaction can be well
incorporated into the model calculation by resorting to an effective 
range expansion fitted to the scattering amplitude of the $\eta N$ model. 
Guided by this finding, we
considered $\eta N$ final state interactions given by effective range expansions with
different values for the scattering length and conclude that the
presently available cross section measurement for the reaction $\gamma d
\rightarrow np\eta$ favor moderate values of the real part of the scattering 
length $a_{\eta N}$.

\vfill

\section*{Acknowledgments}
This work was performed in part under the auspices of the 
U.~S. Department of Energy under contract No. DE-FG02-93ER40756 with 
the Ohio University.  The authors appreciate valuable discussions with
 B.~Krusche, V.~Metag and H.~Str\"oher on the subject.
One of the authors (S.K.) wants to thank the Research Center
for the Subatomic Structure of Matter (CSSM) at the University of Adelaide for 
the warm hospitality during a recent visit,
and acknowledges support by grant No. 447AUS113/14/0 by the Deutsche
Forschungsgemeinschaft.


\begin{table}
\caption{Compilation of different values for the
$\eta{N}$ scattering length $a_{\eta{N}}$
as evaluated by resonance models (RM), T- or K-matrix approaches
or chiral effective Lagrangians ($\chi$EL). The reference in the table
shows only the first author together with the year of publication.
The channels included in the analyses are also listed.} 
\label{tab1}
\begin{tabular}{llccc}
 $a_{\eta{N}}$ (fm) & Reference & Year & Model & Channels \\
\hline
0.27 + i0.22 & Bhalerao~\cite{Bhalerao} & 1985 & RM &$\pi{N}{\to}\pi{N},
\eta{N},\pi\Delta$ \\
0.25 +i0.16 & Bennhold~\cite{Bennhold} & 1991 &  RM &
$\pi{N}{\to}\pi{N},
\eta{N},$ \\
& & & & ${\to}\pi\pi{N};\, \gamma{N}{\to}\eta{N}$\\
0.98 + i0.37 & Arima~\cite{Arima} & 1992 & T & $\pi{N}{\to}\pi{N},
\eta{N}$\\
0.55 +i0.30 & Wilkin~\cite{Wilkin1} & 1993 & FSI & $\pi{N}{\to}\eta{N}$ \\
0.51 +i021 & Sauermann~\cite{Sauerman1}& 1995 &
K & $\pi{N}{\to}\pi{N},
\eta{N}$\\
0.68 + i0.24 & Kaiser~\cite{Kaiser} & 1995 & $\chi$EL &
$\pi{N}{\to}\pi{N},
\eta{N},$ \\
& & & & ${\to}K\Lambda,K\Sigma$ \\
0.888+i0.279 & Batini\'c~\cite{Batinic1} & 1995 &
T & $\pi{N}{\to}\pi{N}, \eta{N}$\\
0.476 +i0.279 & F\"aldt~\cite{Faldt} & 1995 & RM & $\pi{N}{\to}\eta{N}$\\
&&&& $\gamma{N}{\to}\eta{N}$ \\
0.621+i0.306 & Abaev~\cite{Abaev} & 1996 & T &
$\pi{N}{\to}\eta{N}$\\
&&&& ${\bar{K}}N{\to}\eta{\Lambda}$ \\
0.51 +i021 &  Deutsch-Sauermann~\cite{Sauerman2}& 1997 &
K & $\pi{N}{\to}\pi{N},
\eta{N}$ \\
 &   &  &  &  $\gamma{N}{\to}\pi{N},\eta{N}$ \\
0.20 + i0.26 & Kaiser~\cite{Kaiser2} & 1997 & $\chi$EL &
$\pi{N}{\to}\pi{N},
\eta{N},$ \\
0.74 + i0.27 & Green~\cite{Green2} & 1997 & K &
$\pi{N}{\to}\pi{N}, \eta{N}$ \\
& & & & $\gamma{N}{\to}\pi{N},\eta{N}$ \\
0.717+i0.263 &  Batini\'c~\cite{Batinic2} & 1998 & 
T & $\pi{N}{\to}\pi{N}, \eta{N}$\\
0.87 + i0.27 & Green~\cite{Green2} & 1999 & K &
$\pi{N}{\to}\pi{N}, \eta{N}$ \\
& & & & $\gamma{N}{\to}\pi{N},\eta{N}$ \\
1.05+i0.27 & Green~\cite{Green3} & 1999 & K &
$\pi{N}{\to}\pi{N}, \eta{N}$  \\
& & & & $\gamma{N}{\to}\pi{N},\eta{N}$ \\
0.32 + i0.25 & Caro Ramon~\cite{Kaiser3} & 2000 & $\chi$EL &
$\pi{N}{\to}\pi{N}, \eta{N},$ \\
0.772+i0.217 & Nieves~\cite{Nieves} & 2001 & $\chi$EL &
$\pi N \rightarrow \pi N,\eta N,K\Lambda,K\Sigma$ \\
0.54+i0.49 & Krippa~\cite{Krippa} & 2001 & $\chi$EL & $\pi N \rightarrow \pi
N,\eta N$ \\
0.42+i0.34 & present work (Krehl~\cite{Krehl}) & 2001 &  T  
& $\pi N \rightarrow \pi N,\eta N$, etc. \\
\end{tabular}
\end{table}


\newpage

\begin{figure}
\caption{The Feynman diagrams of the meson-baryon model 
 that constitute the Born terms for the transitions in the
$\pi{N}$-${\eta}N$ and the ${\eta}N$-${\eta}N$ channels.}
\label{diag0}
\end{figure}

\begin{figure}
\caption{The real and imaginary parts of the $\eta{N}$ scattering 
amplitude $f$ in the $S_{11}$ partial wave as function of the c.m. 
momentum $q$. The symbols show the results of the J\"ulich meson-baryon 
model \protect\cite{Krehl}. The circles stand for the real part, 
while the squares indicate the imaginary part. The solid lines show 
the effective range expansion using the parameters of Eq.~\protect\ref{ouref}. 
The dashed lines indicate the effective range expansion  with parameters of 
Eq.~\protect\ref{greenef} from Ref.~\protect\cite{Green3}.}
\label{maeta1}
\end{figure}

\begin{figure}
\caption{The real and imaginary parts of the $\eta{N}$  scattering amplitude
$f$ in the  $S_{11}$ partial wave
as function of the c.m.  momentum $q$ showing the the contribution from
the $N^\star$(1535) resonance of the J\"ulich meson-baryon model \protect\cite{Krehl}.
The
circles stand for the real part of $f$, while the squares show the 
imaginary part. As a guide to the eye, the solid lines indicate 
the effective range expansion 
given by the full model, as is shown by the  Fig.~1.}
\label{maeta1s}
\end{figure}

\begin{figure}
\caption{The diagrams for the reaction $\gamma{d}{\to}np\eta$:
(a) impulse approximation (IA), (b) IA with NN final state interaction represented
by the NN t-matrix,
and (c) IA and $\eta{N}$ final state interaction. Here $A$ denotes the 
$\gamma{N}{-}\eta{N}$ transition operator. 
}
\label{diag1}
\end{figure}

\begin{figure}
\caption{The total cross section for inclusive photoproduction of
$\eta$-mesons off deuterium as function of the photon energy $E_\gamma$ 
(lower axis) and the excess energy $\epsilon$ (upper axis).
The experimental data are taken from Ref.~\protect\cite{Krusche1}.
The dotted line shows our IA calculation~\protect\cite{Our}, while the dashed 
line is the result with $np$ final state interaction. The solid line 
shows the full result, including the $\eta{N}$ final state interaction  from the 
J\"ulich meson-baryon model.} 
\label{maeta10_bl}
\end{figure}

\begin{figure}
\caption{The cross section rations for inclusive photon production of $\eta$-mesons
off the deuteron as a function of the excess energy $\epsilon$.
Shown are our calculations including the indicated final state interactions divided by
our calculation~\protect\cite{Our} based on the impulse approximation (dotted line in
Fig.~6). The solid line indicates the full calculation containing $np$ and $\eta{N}$
FSI.  The dashed line stands for a calculation including only the $np$ FSI, whereas the
calculation for the dotted line includes only the $\eta{N}$ FSI.}
\label{maeta3_1}
\end{figure}

\begin{figure}
\caption{The cross section rations for inclusive photon production of $\eta$-mesons
off the deuteron as a function of the excess energy $\epsilon$.
Shown is the experimental cross section divided by our 
calculation~\protect\cite{Our} containing IA and $np$ FSI (dashed line 
of Fig.~6). The solid line indicates the
full calculations containing the $\eta{N}$ FSI divided by the calculation
containing IA and $np$ FSI. The dash-dotted and dashed lines show the 
calculations with $\eta{N}$ scattering lengths 
$a_{\eta{N}}{=}0.74{+}i0.27$~fm and $a_{\eta{N}}{=}0.25{+}i0.16$~fm,
respectively. The hatched area indicate the uncertainty of the
calculations with the $\eta{N}$ FSI taken from the 
J\"ulich meson-baryon model, as discussed in the text.}
\label{maeta3_bl}
\end{figure}

\newpage
\begin{figure}
\begin{center}
\psfig{file=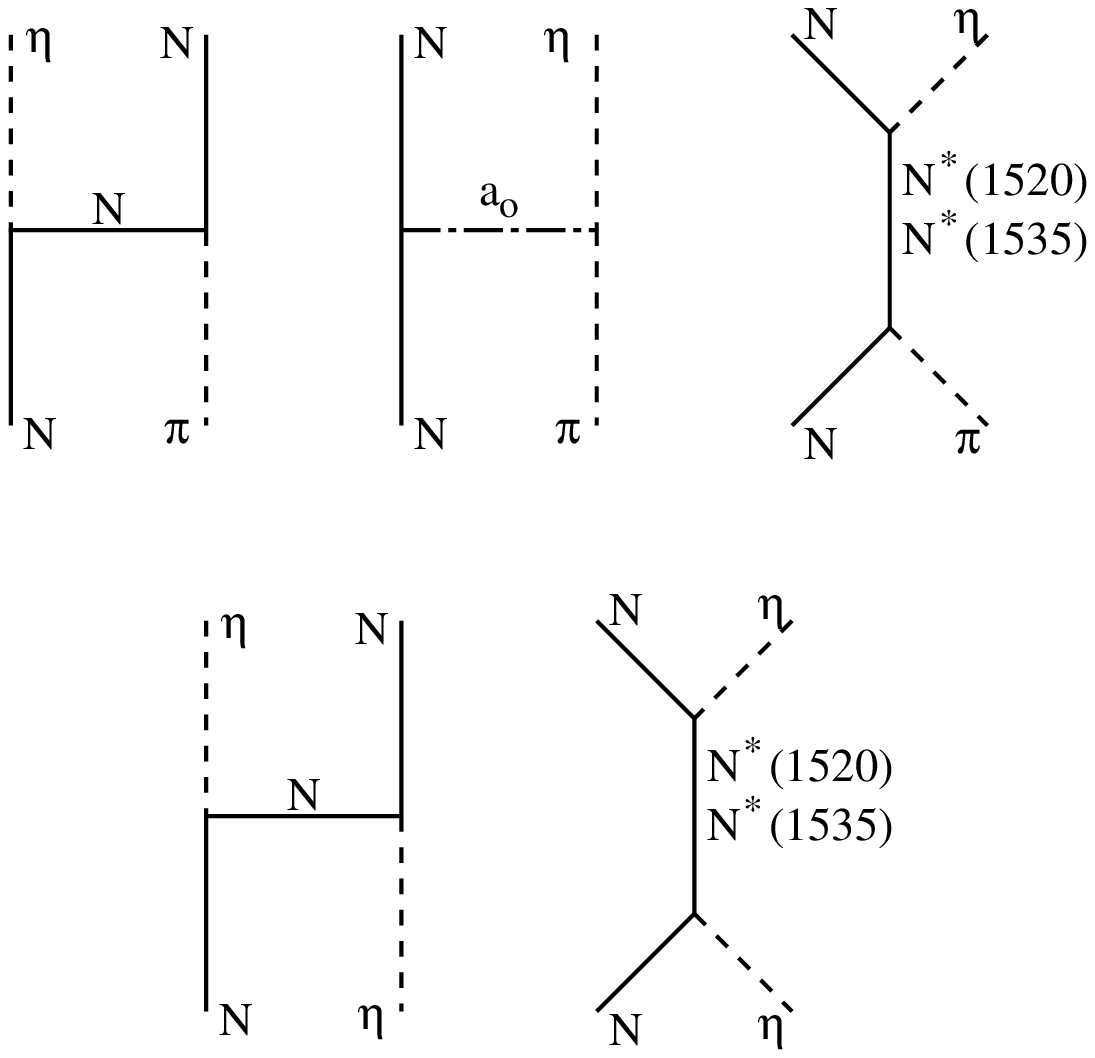,width=10cm,height=9cm}
\vspace{4mm}
\center{FIG. 1}
\end{center}
\end{figure}

\begin{figure}
\begin{center}
\psfig{file=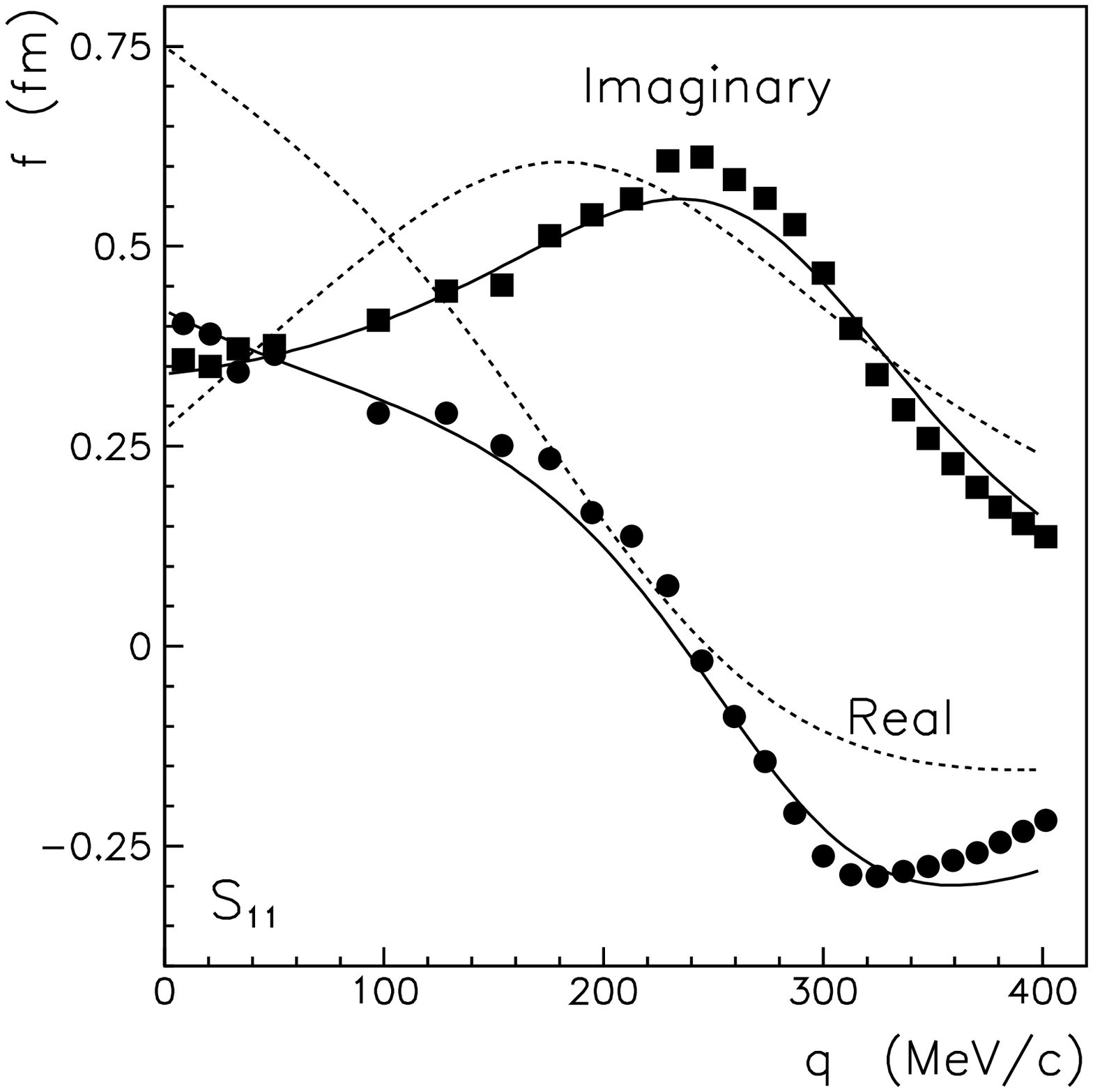,width=10cm,height=9cm}
\vspace*{-5mm}
\center{FIG. 2}
\end{center}
\end{figure}
\newpage

\vspace*{-1cm}
\begin{figure}
\begin{center}
\psfig{file=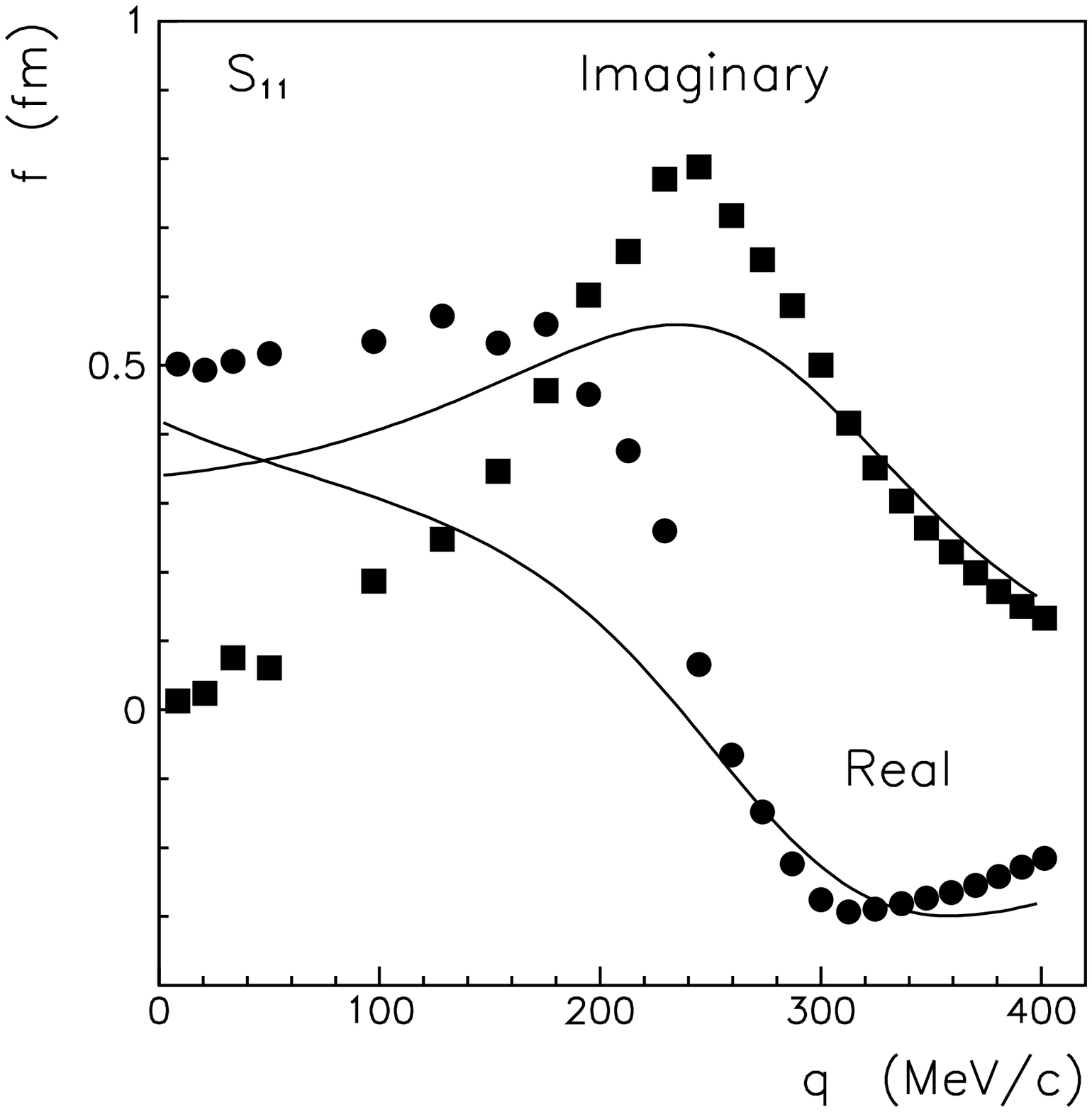,width=10cm,height=9cm}
\vspace*{-5mm}
\center{FIG. 3}
\end{center}
\end{figure}

\begin{figure}
\begin{center}
\psfig{file=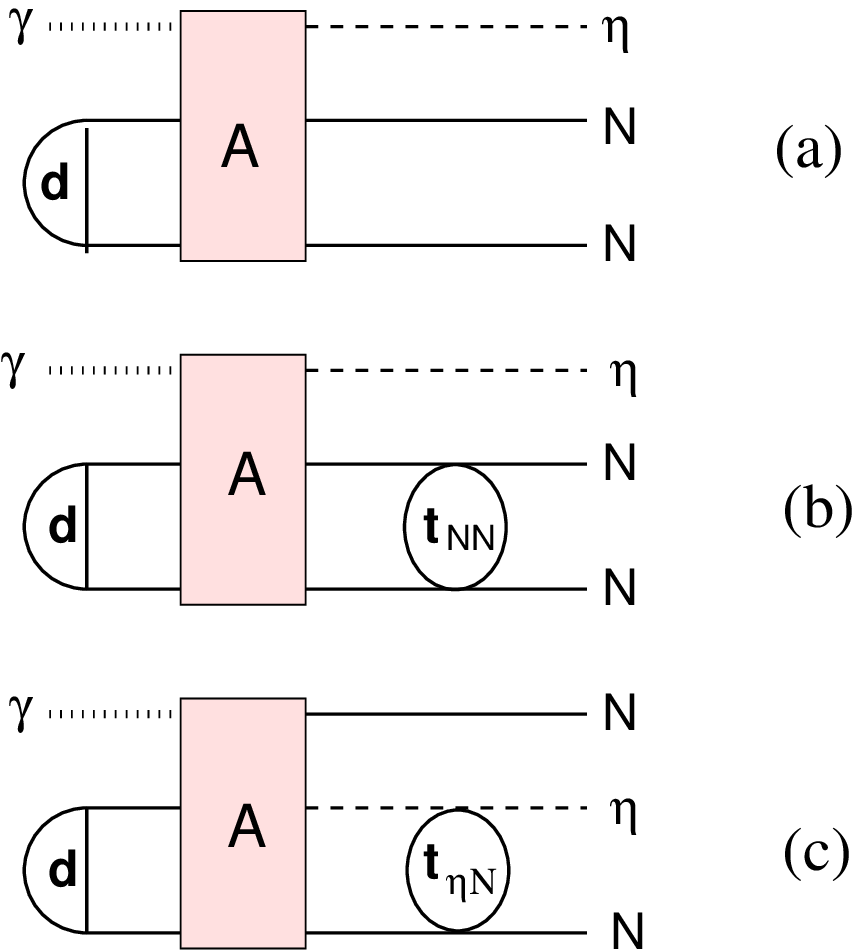,width=7cm}
\vspace*{4mm}
\center{FIG. 4}
\end{center}
\end{figure}
 
\begin{figure}
\begin{center}
\psfig{file=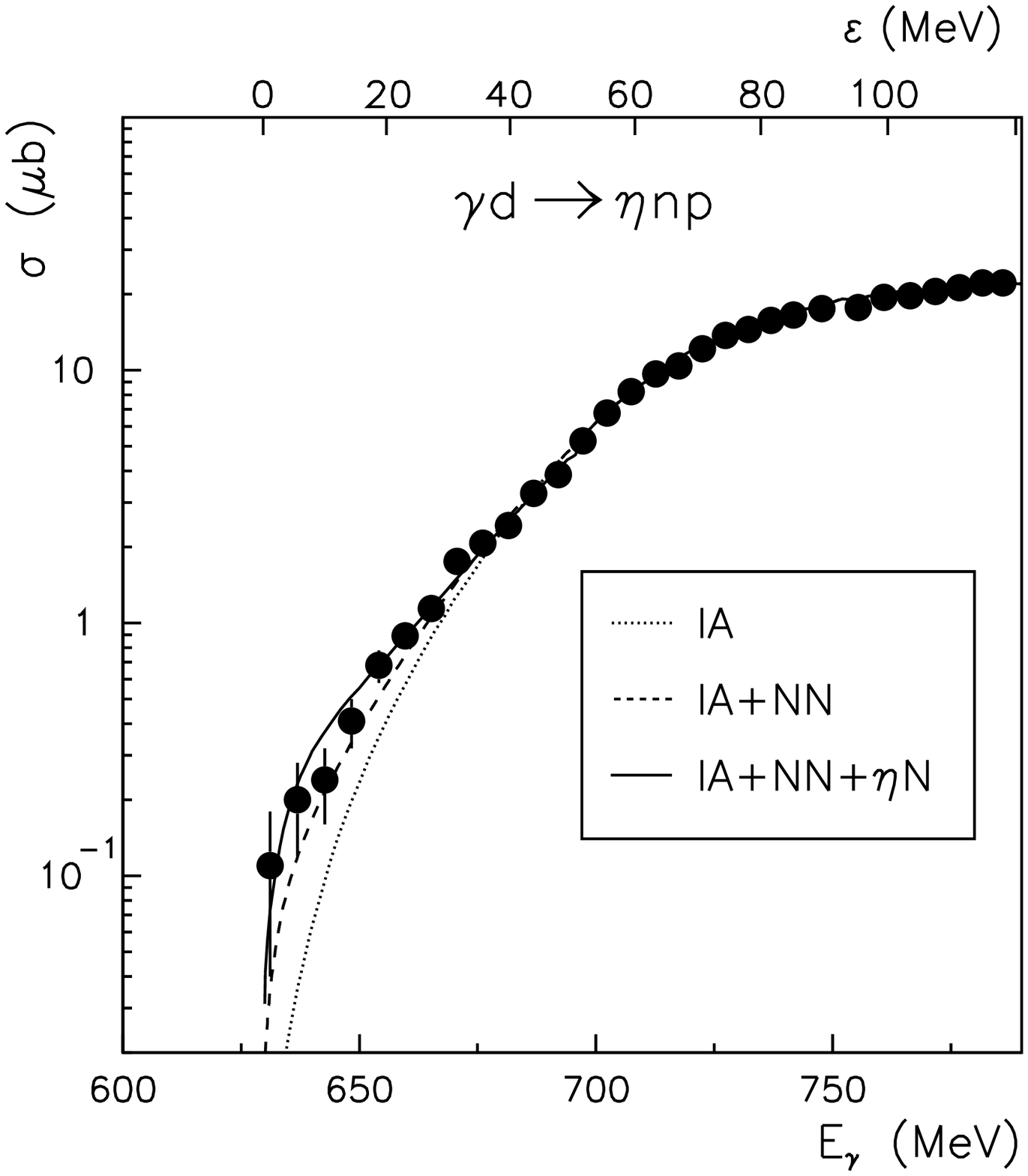,width=10cm,height=9.5cm}
\vspace*{-5mm}
\center{FIG. 5}
\end{center}
\end{figure}

\begin{figure}
\begin{center}
\psfig{file=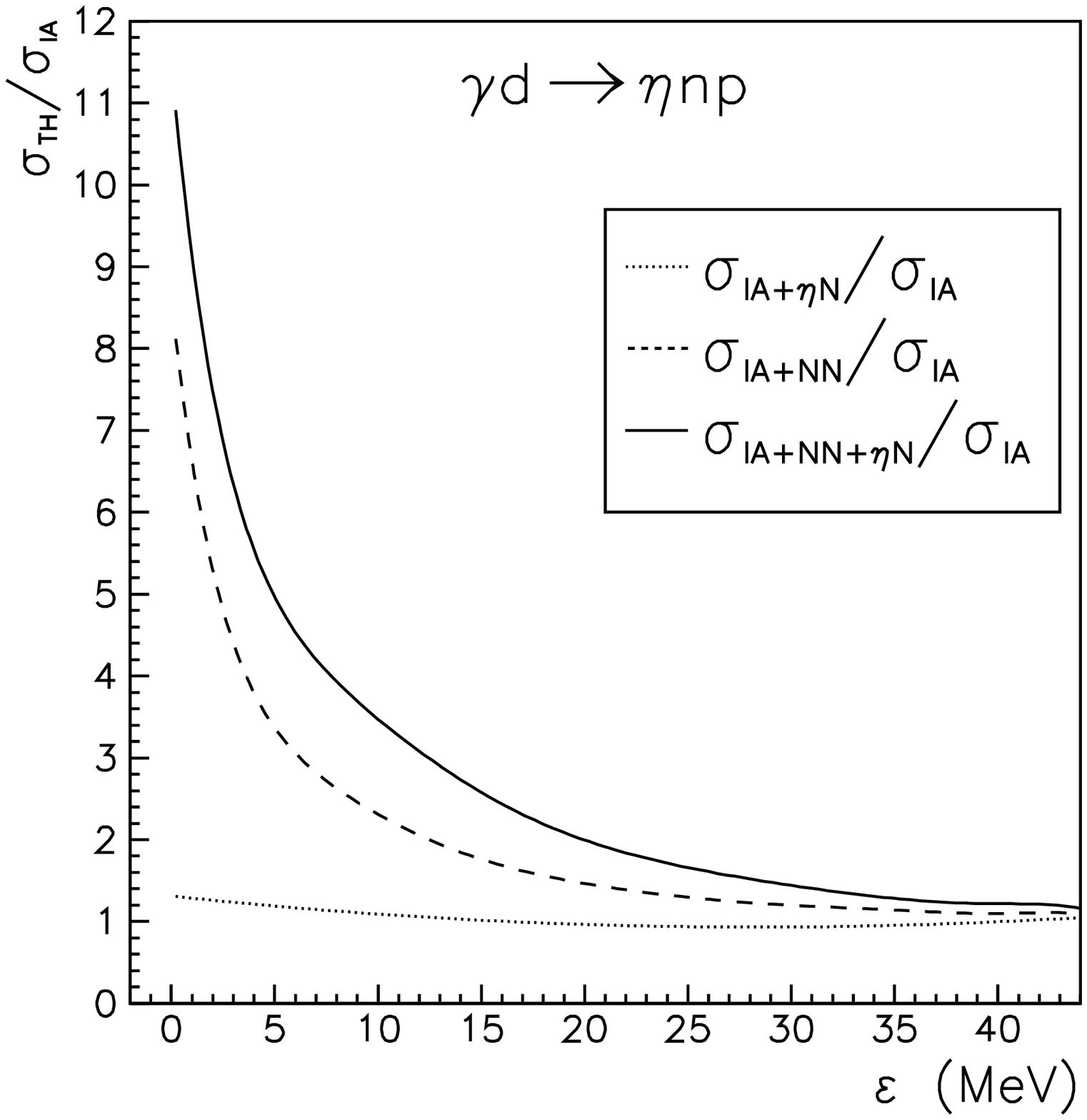,width=10cm,height=9cm}
\vspace*{-5mm}
\center{FIG. 6}
\end{center}
\end{figure}

\begin{figure}
\begin{center}
\psfig{file=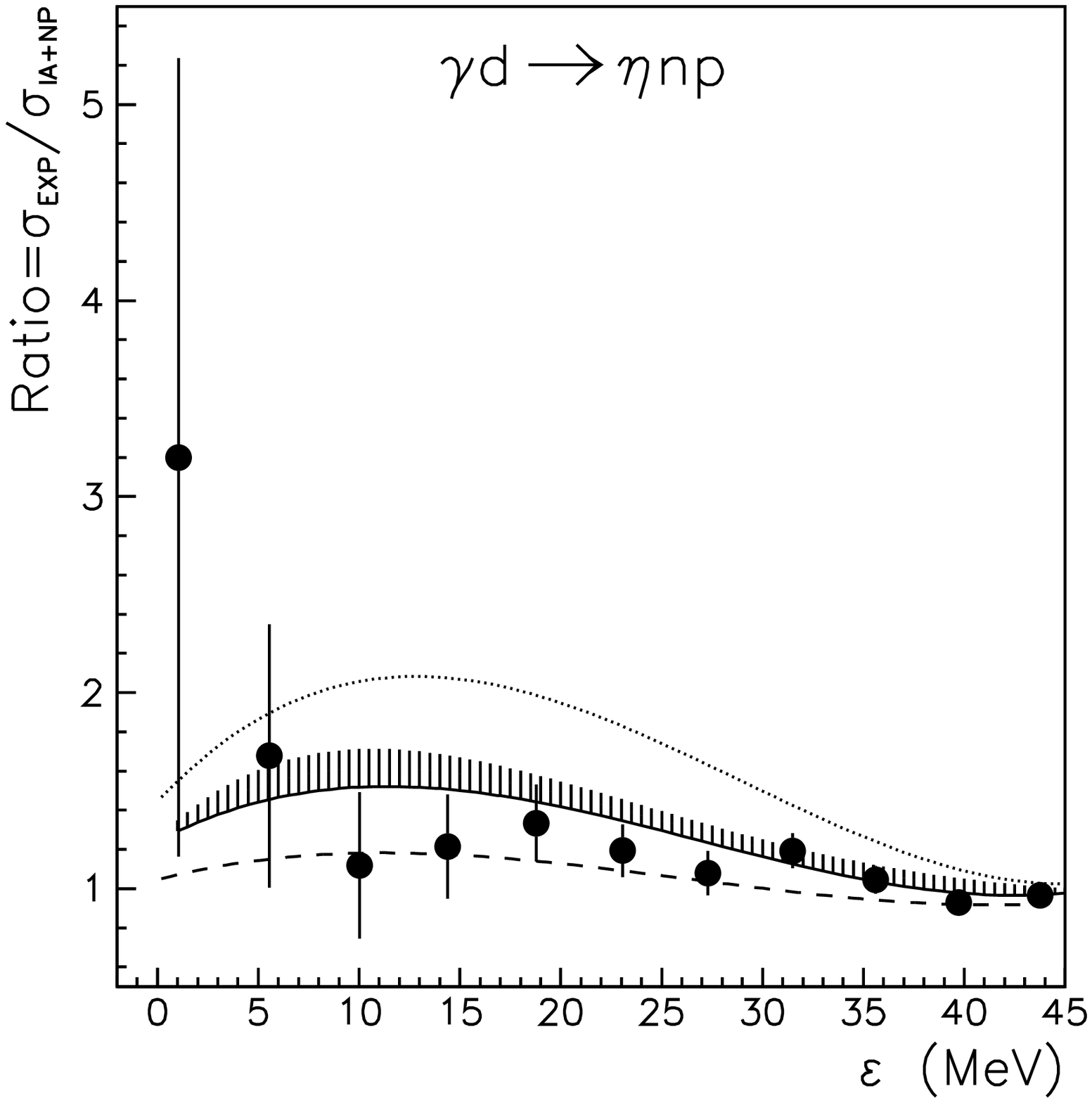,width=10cm,height=9cm}
\vspace*{-5mm}
\center{FIG. 7}
\end{center}
\end{figure}


\begin{references}
\bibitem{Bhalerao}
        R.S. Bhalerao and L.C. Liu, Phys. Rev. Lett. {\bf 54}, 865 (1985).
\bibitem{Bennhold}
        C. Bennhold and H. Tanabe, Nucl. Phys. A {\bf 530}, 625 (1991).
\bibitem{Wilkin1}
        C. Wilkin, Phys. Rev. C {\bf 47}, R938 (1993).
\bibitem{Faldt}
        G. F\"aldt and C. Wilkin, Nucl. Phys. A {\bf 587}, 769 (1995).
\bibitem{Sauerman1}
        Ch. Sauermann, B.L. Friman and W. N\"orenberg, Phys. Lett.
        B {\bf 341}, 261 (1995).
\bibitem{Sauerman2}
        Ch. Deutsch-Sauermann, B.L. Friman and W. N\"orenberg, Phys. Lett.
        B {\bf 409}, 51 (1997).
\bibitem{Green2}
        A.M. Green and S. Wycech, Phys. Rev. C {\bf 55}, R2167 (1997).
\bibitem{Green3}
        A.M. Green and S. Wycech, Phys. Rev. C {\bf 60}, 035208 (1999).
\bibitem{Arima}
        M. Arima, K. Shimizu and K. Yazaki, Nucl. Phys. A {\bf 543}, 
        613 (1992).
\bibitem{Batinic1}
        M.~Batini\'c, I.~Slaus, A.~Svark, B.M.K. Nefkens , Phys. Rev. C {\bf 51}, 2310 (1995);
        M.~Batini\'c, I.~Slaus, A.~Svark, Phys. Rev. C {\bf 52}, 2188 (1995).
\bibitem{Abaev}
        V.V. Abaev and B.M.K. Nefkens, Phys. Rev. C {\bf 53}, 385 (1996).
\bibitem{Batinic2}
    M. Batini\'c, I. Slaus, A. Svark, B.M.K. Nefkens, T.S.H. Lee, Physica Scripta {\bf 58}, 15 (1998).
\bibitem{Kaiser}
        N. Kaiser, P.B. Siegel and W. Weise, Phys. Lett. B {\bf 362}, 
        23 (1995).
\bibitem{Kaiser2}
        N. Kaiser, T. Waas and W. Weise, Nucl. Phys. A {\bf 612}, 
        297 (1997).
\bibitem{Kaiser3}
        J. Caro Ramon, N. Kaiser, S. Wetzel and W. Weise, Nucl. Phys. A {\bf 672}, 
        249 (2000).
\bibitem{Nieves}
        J. Nieves and E. Ruiz Arriola, Phys. Rev. D {\bf 64}, 116008 (2001).
\bibitem{Krippa}
        B. Krippa, Phys. Rev. C {\bf 64}, 047602 (2001). 
\bibitem{Bauer}
        T.H. Bauer, R.D. Spital and D.R. Yennie, Rev. Mod. Phys., {\bf 50},
        261 (1978)
\bibitem{Effenberger}
        M. Effenberger and A. Sibirtsev, Nucl. Phys. A {\bf 632}, 99 (1998).
\bibitem{Mayer1}
        B. Mayer et al., Phys. Rev. C {\bf 53}, 2068 (1996).
\bibitem{Betigeri}
         M. Betigeri et al., Phys. Lett. B {\bf 472},  267 (2000). 
\bibitem{Klimala}
        W. Klimala et al., Acta Phys.\ Polon.\ B {\bf 31}, 2231 (2000). 
\bibitem{Calen1}
        H. Cal\'en et al., Phys. Rev. Lett. {\bf 79}, 2642 (1997).
\bibitem{Calen2}
        H. Cal\'en et al., Phys. Rev. Lett. {\bf 80}, 2069 (1998).
\bibitem{Chiavassa}
        E. Chiavassa et al., Phys. Lett. B {\bf 322}, 270 (1994).
\bibitem{Calen3}
        H. Cal\'en et al., Phys. Lett. B {\bf 366}, 39 (1996).
\bibitem{Bergold}
        A.M. Bergdolt et al., Phys. Rev. D {\bf 48}, R2969 (1993).
\bibitem{Smyrski}
        J. Smyrski et al., Phys. Lett. B {\bf 474}, 182 (2000).
\bibitem{Krusche1}
        B. Krusche et al., Phys. Lett. B \textbf{358}, 40 (1995).
\bibitem{Our}
        A. Sibirtsev, Ch. Elster, J. Haidenbauer and J.~Speth,
        Phys. Rev. C {\bf 64}, 024006.
\bibitem{Schutz99}
        C. Sch\"utz, J. Haidenbauer, J. Speth, and J.W. Durso,
 Phys. Rev. C {\bf 57}, 1464 (1998).
\bibitem{Krehl}
        O. Krehl, C. Hanhart, S. Krewald, J. Speth, Phys. Rev. C {\bf 62},
        025207 (2000).
\bibitem{Janssen}
        G. Janssen, B.C. Pearce, K. Holinde, and J. Speth, Phys. Rev. D {\bf 52},
        2690 (1995).
\bibitem{Feuster} T. Feuster and U. Mosel, Phys. Rev. C {\bf 59}, 460(1999).
\bibitem{Machleidt1} 
        R. Machleidt, Phys. Rev. C \textbf{63}, 024001 (2001).
\bibitem{Green1}
        A.M. Green and  S. Wycech, Phys. Rev. C {\bf 55}, 2167 (1997).
\end{references}
\end{document}